\documentclass[fleqn]{ILASS_2017_Paper_Format}

\usepackage{epstopdf}
\usepackage[per-mode=symbol]{siunitx}
\usepackage[tight]{units}
\usepackage{color}

\usepackage[section]{placeins}

\usepackage[protrusion=true,  %% enable prostrusion
expansion,      %% enable expansion
kerning=true,   %% enable kerning
tracking=true,  %% enable tracking -> affects sc text!
final           %% enable microtype (use 'draft' to disable)
]{microtype}

\newcommand{\myunitbr}[2]{%
  \unit[{#1}]{{[#2]}}%
}%

\DeclareSIUnit\pixel{px}
\newcommand{\Web}{\mbox{\textit{We}}}   % Weber number

\begin{document}

% ------------------------------------------------------------------------------
% ------------------------------------------------------------- title page -----

\title{Drop Stream - Immiscible Jet Collisions:\\Regimes and Fragmentation Mechanisms}

\doi{http://dx.doi.org/10.4995/ILASS2017.2017.4707}
\firstpagehead{ILASS--Europe 2017, 28th Conference on Liquid Atomization and Spray Systems, 6-8 September 2017, Valencia, Spain}
\runningheads{ILASS -- Europe 2017, 6-8 Sep. 2017, Valencia, Spain}{}
\runnigfoots{This work is licensed under a Creative Commons 4.0 International License (CC BY-NC-ND 4.0). \\ EDITORIAL UNIVERSITAT POLITÈCNICA DE VALÈNCIA
}

\author{Carole Planchette$^{*}$, Hannes Hinterbichler, Günter Brenn}

\address{Institute of Fluid Mechanics and Heat Transfer, Graz University of Technology, Austria\\
\vskip 1pt
{\normalsize{*Corresponding author:
\href{mailto:carole.planchette@tugraz.at}{\textstyleInternetlink{carole.planchette@tugraz.at}}}}}
\vskip 8 mm

% ------------------------------------------------------------------------------
% --------------------------------------------------------------- abstract -----

\section{Abstract}
We investigate the collision of a continuous liquid jet with a regular stream of immiscible droplets.
The immiscible liquids, namely silicon oil for the continuous jet and an aqueous glycerol solution for the drop stream, are selected to enable the total wetting of the drops by the jet liquid. 
Four different regimes are experimentally identified: \textit{drops in jet}, \textit{encapsulation without satellites}, \textit{encapsulation with satellites from the jet liquid} and \textit{mixed fragmentation}.
The \textit{drops in jet} regime, potentially of great interest for new applications, corresponds to a regular stream of drops embedded in a continuous jet and is described and reported for the first time.
Using well known aspects of drop collision and jet stability, we propose to model the transition between the \textit{drops in jet} regime and the others. Two dimensionless parameters are derived from this analysis which are thus used to produce a simple regime map where the \textit{drops in jet} regime can be well distinguished from the other outcomes.

\section{Keywords}
drops, jet, breakup/fragmentation, collisions, capillary flows

% ------------------------------------------------------------------------------
% ----------------------------------------------------------- introduction -----

\section{Introduction}
Interactions and stability of fluid systems with free surfaces have been of scientific interest since more than a century due to their importance for processes in nature and industry. Collisions between drops have been widely studied, involving, for example, two identical drops \cite{ref:Jiang-Umemura-Law_1992}, two drops of immiscible liquids \cite{ref:Planchette-Lorenceau-Brenn_2012}, or even three drops that come simultaneously into contact \cite{ref:Hinterbichler-Planchette-Brenn_2015}. Liquid jets and ligaments, which may among others originate from drop collisions and impacts, have also been deeply investigated \cite{ ref:Villermaux_2007,ref:Eggers_2008}. Their fragmentation, as well as the collisions between two jets \cite{ref:Bremond_2006}, have been in focus since they may result in the production of drops which must be occasionally avoided or enhanced and controlled for processes such as fiber production, atomization, printing, etc.

In the present study, the collisions of a drop stream with a continuous jet consisting of different immiscible liquids are experimentally investigated. Special interest is given to the possibility to incorporate a regular drop stream into a continuous liquid jet without causing its fragmentation. Such a structure may lead to promising applications, especially in the field of encapsulation. We refer to it as the \textit{drops in jet} regime and focus on its limit of stability.

In the following section we present our materials and methods, including the experimental setup.
Our results are presented thereafter. The observed regimes are described, and we propose an analysis to model the transition between the \textit{drops in jet} regime and the others. The paper ends with the conclusions.

% ------------------------------------------------------------------------------
% -------------------------------------------------- materials and methods -----

\section{Materials and methods}

The jet and the drops (as well as the associated liquids) are designated by the subscripts $j$ and $d$, respectively. The full description of the collisions requires knowing the fluid properties of the immiscible liquids, as well as some geometric and kinetic parameters.

% ----------------------------------------------------- immiscible liquids -----

\subsection{Immiscible liquids}
We use an aqueous glycerol solution (\SI{50}{\percent} glycerol by weight) for the drop stream and a silicon oil of low viscosity for the jet (silicon oil M5, from Carl Roth, Austria). The aqueous glycerol solution is colored with food dye to distinguish it from the silicon oil on the images.

The relevant physical properties of these liquids, density $\rho$, dynamic viscosity $\mu$ and the surface tension $\sigma$ are listed in table~\ref{tab:fluid_properties}. The interfacial tension between them, denoted $\sigma_{dj}$, is equal to 35 $\si{\milli\newton\per\metre}$.
Note that, due to the relative values of the surface and interfacial tensions, the total wetting of the drops by the jet liquid is ensured.

\begin{table}[h]
\centering
\caption{Physical properties of the used liquids at $23\pm\SI[mode=math]{2}{\celsius}$. The glycerol content is given in mass percent.}
\vspace{6pt}
\begin{tabular}{l S S S}
\toprule
{Liquid} & {Density} & {Dynamic visocity} & {Surface tension} \\
& $\myunitbr{\rho}{\si{\kilogram\per\cubic\metre}}$
& $\myunitbr{\mu}{\si{\milli\pascal\second}}$
& $\myunitbr{\sigma}{\si{\milli\newton\per\metre}}$ \\
\midrule 
Glycerol \SI{50}{\percent} (G50)   & 1131.3 & 5.24 & 66.53 \\
Silicon oil m5 (SO M5)             &  913.4 & 4.57 & 19.50 \\
\bottomrule
\end{tabular}
\label{tab:fluid_properties}
\end{table}

% ----------------------------------------------------- experimental setup -----

\subsection{Experimental setup}

\begin{figure}[h]
\centering
\includegraphics[scale=0.8]{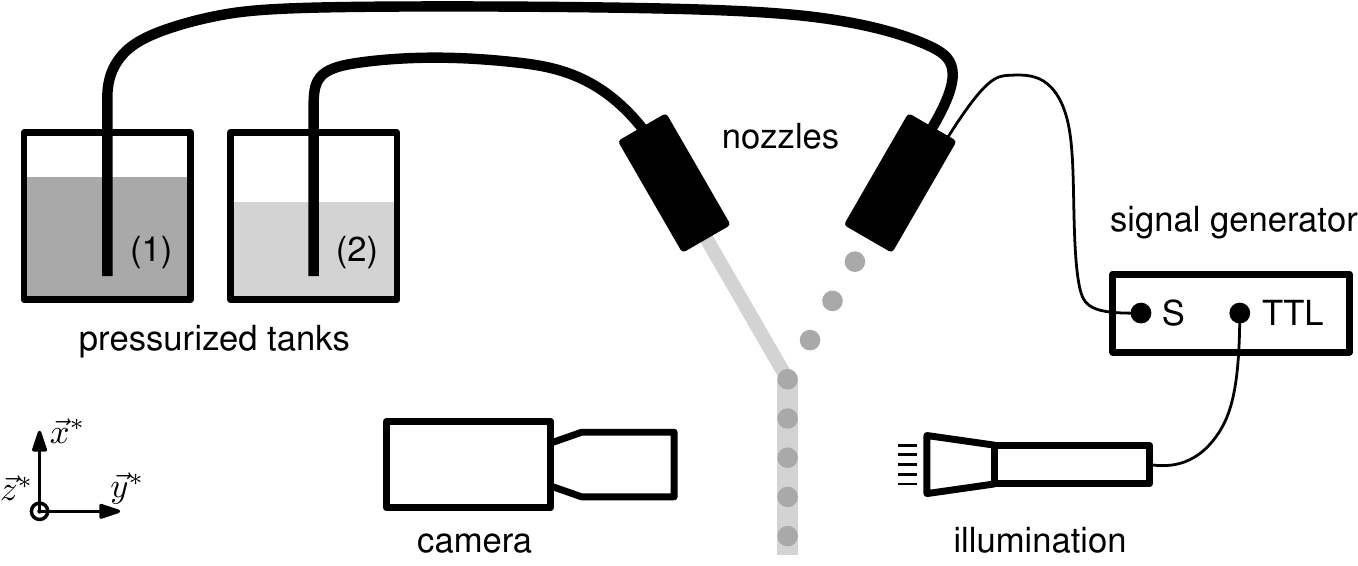}
\caption{Experimental setup for collisions of a droplet stream with a continuous and immiscible liquid jet.}
\label{fig:exp_setup}
\end{figure}

To produce controlled collisions between the continuous jet and the monodisperse droplet stream, the experimental setup of figure~\ref{fig:exp_setup} is used.
It consists of two independent pressurized tanks which supply the immiscible liquids (G50 and SO M5)  allowing for independent flow rate adjustment.
Two nozzles with variable orifice diameter produce liquid jets that can be disturbed at an adjustable frequency via a piezo electric crystal to make use of the Plateau-Rayleigh instability and produce regular droplet streams \cite{ref:Brenn-Durst-Tropea_1996}.
Thus, in this work, one nozzle is connected to a signal generator to provide a regular droplet stream, while the other one is used unconnected to supply a continuous liquid jet.

The typical frequency of drop formation $f$ is in the order of \SI{10}{\kilo\hertz}.
The jet and drop diameters are not necessarily identical and range between \SI{100}{\micro\meter} and \SI{300}{\micro\meter}.
Both nozzles are mounted on micro-stages to allow fine translational and rotational adjustment of the liquid trajectories.
The collisions are illuminated by stroboscopic lighting with the help of an LED lamp.
Pictures are taken with a PCO Sensicam video camera.
The typical resolution of our imaging system is in the order of {\SI{2}{\micro\meter\per\pixel}}.

% --------------------------------------- kinetic and geometric parameters -----

\subsection{Kinetic and geometric parameters}

\begin{figure}[b]
\centering
\includegraphics{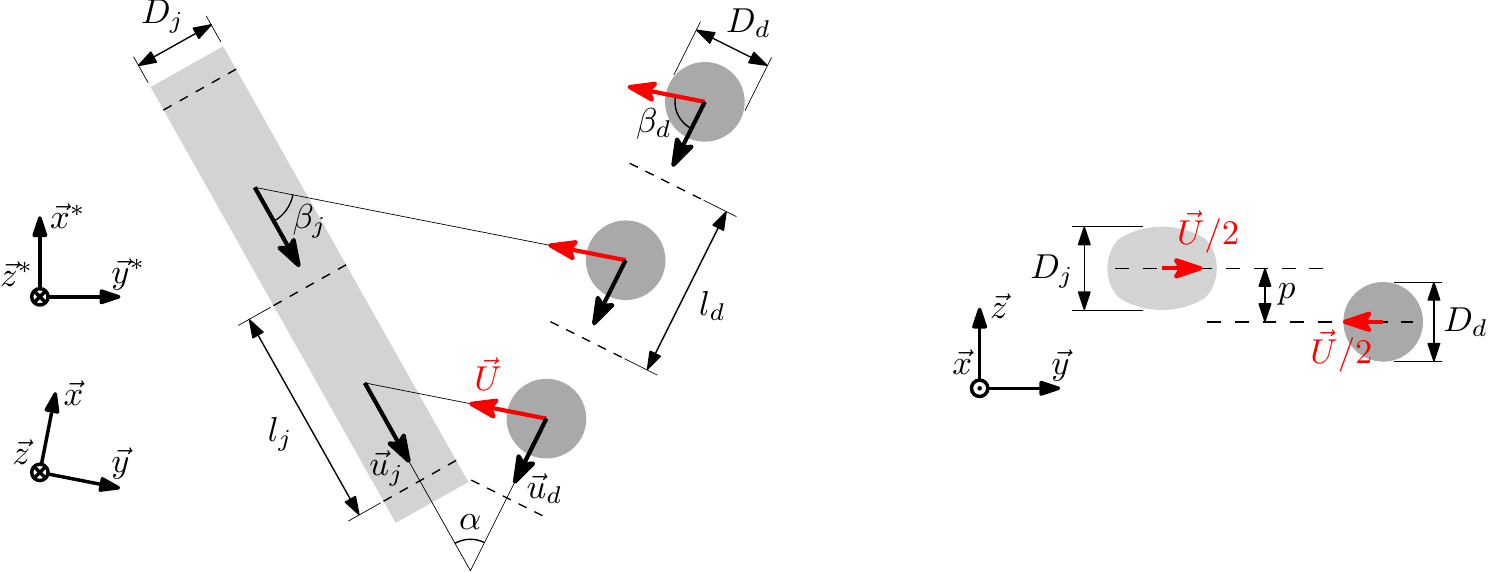}
\caption{Geometric and kinetic parameters of the collision. $(\vec{x}^{\ast}, \vec{y}^{\mspace{1mu}\ast}, \vec{z}^{\mspace{1mu}\ast})$ represents the laboratory reference frame and $(\vec{x}, \vec{y}, \vec{z})$ the reference frame moving with the jet, where $\vec{y}$ is collinear with the relative velocity $\vec{U}$.}
\label{fig:sketch_collision}
\end{figure}

The geometric and kinetic parameters of the studied collisions  are illustrated in figure~\ref{fig:sketch_collision}.
In the laboratory reference frame $(\vec{x}^{\ast}, \vec{y}^{\mspace{1mu}\ast}, \vec{z}^{\mspace{1mu}\ast})$, the jet (diameter $D_{j}$) and the drops (diameter $D_{d}$) have the velocities $\vec{u}_{j}$ and $\vec{u}_{d}$, respectively.
In the reference frame moving with the jet $(\vec{x}, \vec{y}, \vec{z})$, the velocities $\vec{u}_{j}$ and $\vec{u}_{d}$ can be replaced by the relative velocity $\vec{U}$, which is collinear with $\vec{y}$.
$l_{j}$ and $l_{d}$ are the spatial periods, where $l_{d}$ can be tuned varying the flow rate of the stream and the frequency of drop production $f$.
The spatial period of the jet can then by calculated by
\begin{equation}
    \label{eq:spatial_period_jet}
  l_{j} = l_{d}\frac{u_{j}}{u_{d}}\,.
\end{equation}
The impact parameter $p$, see the right hand side of figure~\ref{fig:sketch_collision}, is defined as the distance separating the droplet center of mass and the one of the corresponding volume for the jet (length $l_{j}$) perpendicular to their relative direction of movement.
For the present study, only head-on collisions $(p=0)$ are considered. Practically, the position $p=0$ can be found by observing the evolution of the collision outcome during the translation of one of the nozzles along $\vec{z}^{\mspace{1mu}\ast}$. Detecting the position where this evolution reverses accurately provides $p=0$. Furthermore, the collisions are symmetric with respect to $\vec{x}^{\ast}$ ($\alpha$ is bisected by $\vec{x}^{\ast}$).

$D_{j}$, $D_{d}$, $l_{d}$ and all the relevant angles (see figure~\ref{fig:sketch_collision}) are obtained by image analysis using the public domain software ImageJ \cite{ref:imagej}.
The velocity of the drops is given by $u_{d}=l_{d}\,f$.
The velocity of the jet $u_{j}$ is deduced from continuity, measuring the flow rate of the jet and its diameter.
The relative velocity $\vec{U}$ yields
\begin{equation}
    \label{eq:relative_velocity}
  \vec{U} = \vec{u}_{j} - \vec{u}_{d}
\end{equation}
and can be modified by varying $u_{j}$, $u_{d}$ or $\alpha$.

% ------------------------------------------------------------------------------
% ------------------------------------------------- results and discussion -----

\section{Results and Discussion}

% ---------------------------------------------------------------- regimes -----

\subsection{Regimes}
\begin{figure}[h]
\centering
  \parbox[c]{0.03\textwidth}{(a)}
  \hspace*{0mm}
  \parbox[c]{0.75\textwidth}{\includegraphics[angle=90,width=0.82\textwidth]{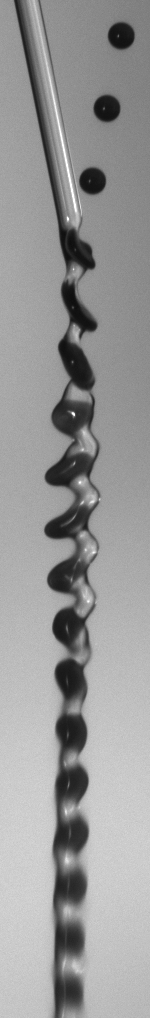}}\\
  \vspace{1.5mm}
  \parbox[c]{0.03\textwidth}{(b)}
  \hspace*{0mm}
  \parbox[c]{0.75\textwidth}{\includegraphics[angle=90,width=0.82\textwidth]{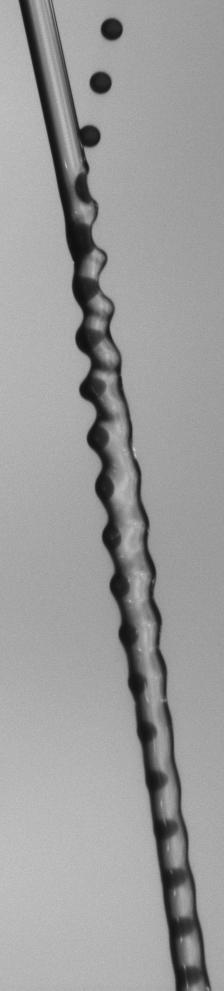}}
  
\caption{Regime \textit{drops in jet}:
  (a), $D_{d}=\SI{202}{\micro\meter}$, $D_{j}=\SI{200}{\micro\meter}$,
    $u_{d}=\SI{6.48}{\meter\per\second}$, $u_{j}=\SI{3.43}{\meter\per\second}$, $U=\SI{3.56}{\meter\per\second}$;
  (b), $D_{d}=\SI{192}{\micro\meter}$, $D_{j}=\SI{300}{\micro\meter}$,
    $u_{d}=\SI{5.55}{\meter\per\second}$, $u_{j}=\SI{4.92}{\meter\per\second}$, $U=\SI{2.30}{\meter\per\second}$.}
    \label{fig:regime_1}
\end{figure}

\begin{figure}[b]
\centering
  \parbox[c]{0.03\textwidth}{(a)}
  \hspace*{0mm}
  \parbox[c]{0.75\textwidth}{\includegraphics[angle=90,width=0.82\textwidth]{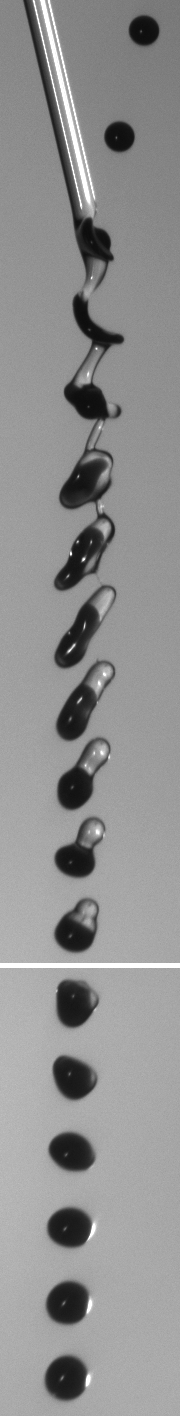}}\\
  \vspace{1.5mm}
  \parbox[c]{0.03\textwidth}{(b)}
  \hspace*{0mm}
  \parbox[c]{0.75\textwidth}{\includegraphics[angle=90,width=0.82\textwidth]{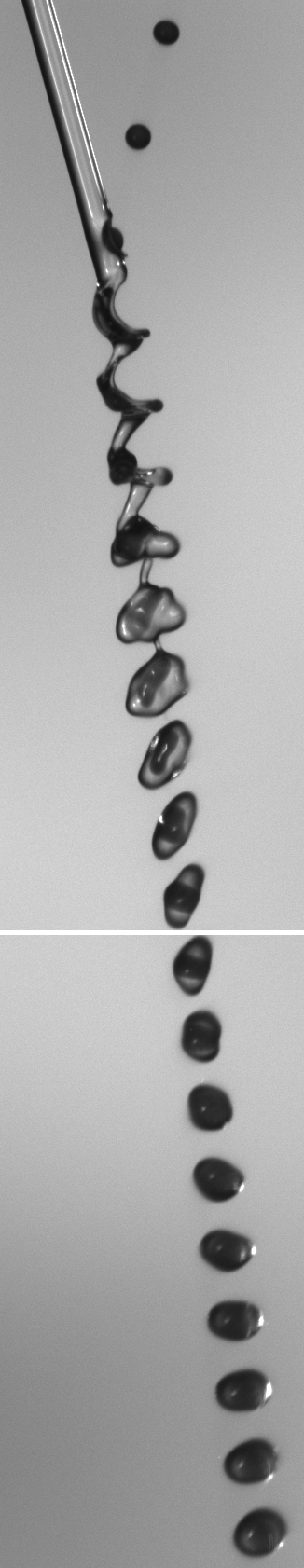}}
  
\caption{Regime \textit{encapsulation without satellites}:
  (a), $D_{d}=\SI{227}{\micro\meter}$, $D_{j}=\SI{200}{\micro\meter}$,
    $u_{d}=\SI{6.56}{\meter\per\second}$, $u_{j}=\SI{3.43}{\meter\per\second}$, $U=\SI{3.80}{\meter\per\second}$;
  (b), $D_{d}=\SI{238}{\micro\meter}$, $D_{j}=\SI{300}{\micro\meter}$,
    $u_{d}=\SI{8.09}{\meter\per\second}$, $u_{j}=\SI{4.93}{\meter\per\second}$, $U=\SI{4.60}{\meter\per\second}$.}
    \label{fig:regime_2}
\end{figure}

\begin{figure}[h]
\centering
  \parbox[c]{0.03\textwidth}{(a)}
  \hspace*{0mm}
  \parbox[c]{0.75\textwidth}{\includegraphics[angle=90,width=0.82\textwidth]{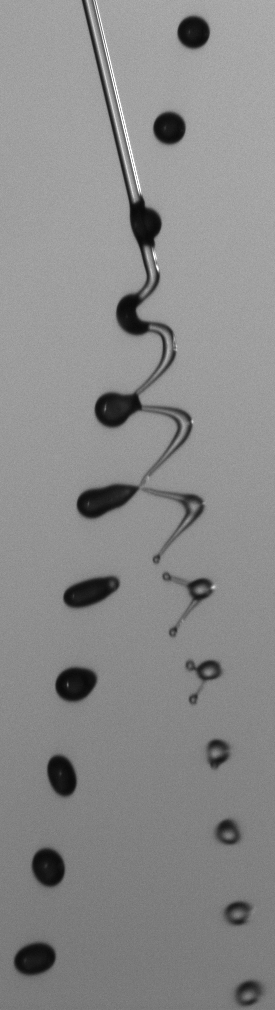}}\\
  \vspace{1.5mm}
  \parbox[c]{0.03\textwidth}{(b)}
  \hspace*{0mm}
  \parbox[c]{0.75\textwidth}{\includegraphics[angle=90,width=0.82\textwidth]{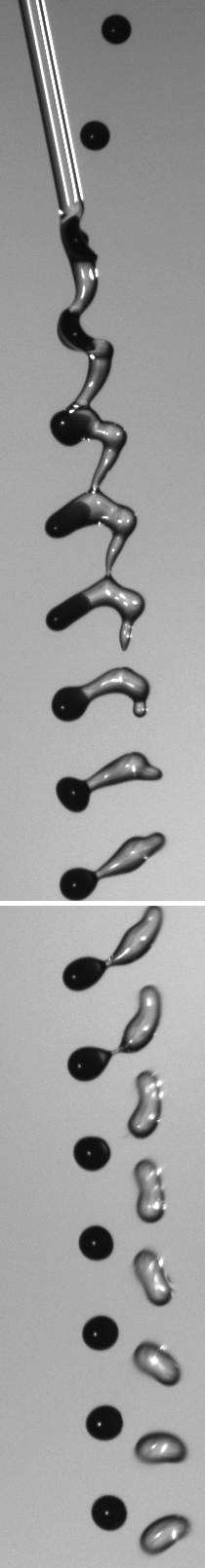}}
  
\caption{Regime \textit{encapsulation with pure oil satellites}:
  (a), $D_{d}=\SI{210}{\micro\meter}$, $D_{j}=\SI{100}{\micro\meter}$,
    $u_{d}=\SI{7.60}{\meter\per\second}$, $u_{j}=\SI{6.03}{\meter\per\second}$, $U=\SI{3.58}{\meter\per\second}$;
  (b), $D_{d}=\SI{229}{\micro\meter}$, $D_{j}=\SI{200}{\micro\meter}$,
    $u_{d}=\SI{6.75}{\meter\per\second}$, $u_{j}=\SI{4.98}{\meter\per\second}$, $U=\SI{2.89}{\meter\per\second}$.}
    \label{fig:regime_3}
\end{figure}

\begin{figure}[!b]
\centering
  \parbox[c]{0.03\textwidth}{(a)}
  \hspace*{0mm}
  \parbox[c]{0.75\textwidth}{\includegraphics[angle=90,width=0.82\textwidth]{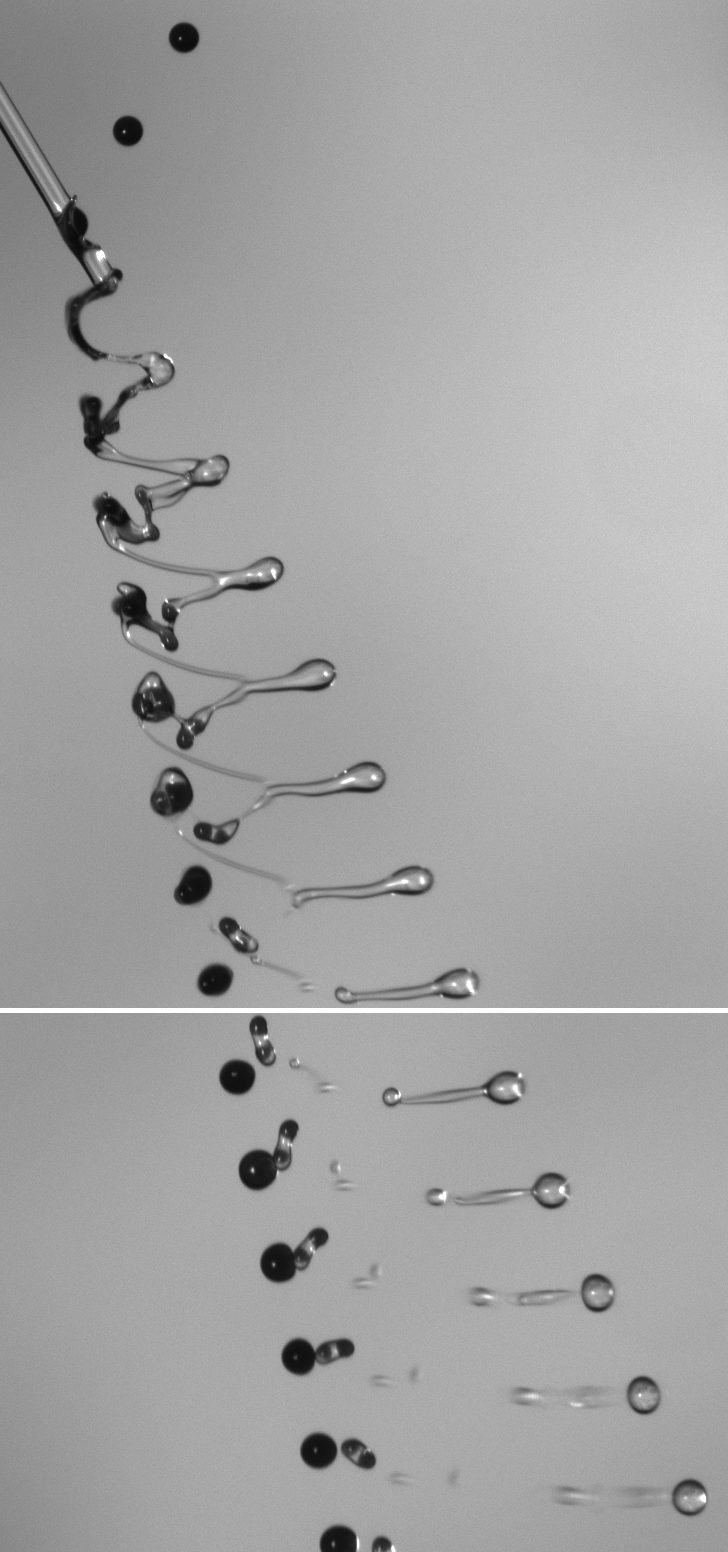}}\\
  \vspace{1.5mm}
  \parbox[c]{0.03\textwidth}{(b)}
  \hspace*{0mm}
  \parbox[c]{0.75\textwidth}{\includegraphics[angle=90,width=0.82\textwidth]{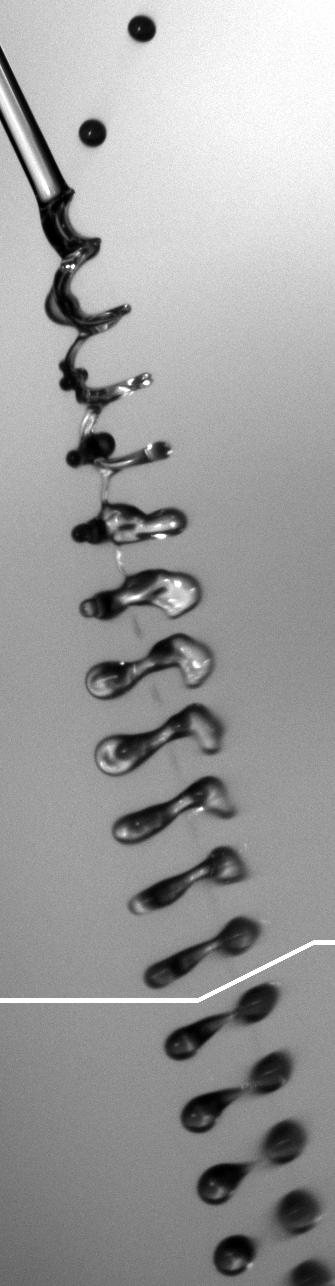}}
  
\caption{Regime \textit{multiple or mixed fragmentation}:
  (a), $D_{d}=\SI{223}{\micro\meter}$, $D_{j}=\SI{200}{\micro\meter}$,
    $u_{d}=\SI{6.60}{\meter\per\second}$, $u_{j}=\SI{7.22}{\meter\per\second}$, $U=\SI{7.12}{\meter\per\second}$;
  (b), $D_{d}=\SI{236}{\micro\meter}$, $D_{j}=\SI{300}{\micro\meter}$,
    $u_{d}=\SI{8.08}{\meter\per\second}$, $u_{j}=\SI{4.93}{\meter\per\second}$, $U=\SI{6.23}{\meter\per\second}$.}
    \label{fig:regime_4}
\end{figure}

We identified four different outcomes for the collision of a droplet stream with an immiscible continuous jet.
These regimes are described below with the help of collision photographs.
In all photographs (figures~\ref{fig:regime_1}-\ref{fig:regime_4}), the droplets and the jet move from left to right. An (almost) horizontal white breaking line in a picture indicates that separate pictures were needed to capture the full process.

\begin{itemize}
\item {Drops in jet}

  The \textit{drops in jet} regime (see figure~\ref{fig:regime_1}) exhibits a regular stream of droplets embedded in an immiscible liquid jet which does not fragment.
In figure~\ref{fig:regime_1}a, the jet and drops have the same diameter of approximately \SI{200}{\micro\meter}.
The two velocities ($u_{d}=\SI{6.48}{\meter\per\second}$, $u_{j}=\SI{3.43}{\meter\per\second}$) represent similar momentum fluxes in the laboratory reference frame, as it can be seen downstream from the direction of movement of the resulting stream.
In figure~\ref{fig:regime_1}b, the jet momentum flux (measured in the laboratory reference frame) is greater than the one of the droplet stream. As a result, the direction of the resulting \textit{drops in jet} structure is almost similar to the one of the jet alone before the collision. 
In both cases, the regularity of the drop spacing is preserved, although its actual value may be modified by the collision.
This effect originates from the relative orientation of the jet and drop velocities, and a geometric analysis leads to a drop spacing after encapsulation equal to $l_j=l_d\sin(\beta_d )/\sin(\beta_j ) =l_d\,u_j/u_d $.
\end{itemize}

\begin{itemize}
\item {Encapsulation without satellites}
    \label{subsec:regime_2}

  In figure~\ref{fig:regime_2}, two examples of \textit{encapsulation without satellites} can be seen.
The drops of the stream are embedded by an immiscible shell made of the liquid from the fragmented jet. The resulting structure is a stream of individual spherical capsules in the surrounding air. No satellites are formed, and each drop initially present in the drop stream gives rise to a unique capsule. Thus, the shell volume corresponds to the volume of a cylinder of diameter $D_{j}$ and length $l_{j}$.
The resulting stream of capsules is very regular, showing a spacing potentially different from the initial drop spacing. Here again, the spacing modification originates from the relative orientation of the jet and drop velocities. As for the \textit{drops in jet} regime, the relative importance of the initial jet and initial drop stream momentum fluxes is visible in the orientation of the final structure, balanced for picture (a) and dominated by the jet for picture (b).

\end{itemize}

\begin{itemize}
\item{Encapsulation with pure oil satellites}
    \label{subsec:regime_3}

Two illustrations of the \textit{encapsulation with pure oil satellites} regime are reproduced in figure~\ref{fig:regime_3}.
The drops cross over the jet which fragments. Its liquid is distributed between the shell of spherical capsules and satellite droplets. The spherical capsules form a regular stream. The core of each capsule has the volume of one drop initially present in the drop stream, while its immiscible shell has a volume corresponding to a portion of the volume of a cylinder of diameter $D_{j}$ and length $l_{j}$. The excess of jet liquid forms a regular stream of satellite drops. 

\end{itemize}

\begin{itemize}
\item{Multiple or mixed fragmentation}
   \label{subsec:regime_4}  

In figure~\ref{fig:regime_4}, the last identified regime, \textit{multiple or mixed fragmentation}, can be observed.
This regime typically occurs at large relative velocities, which can be seen by the large deformations of the jet and drops upon impact.
In contrast to all the other regimes, not only the jet fragments, but also the drops initially present in the stream. As a result, a structure is formed exhibiting several droplets containing both liquids.
Oil satellites may also be formed (see figure~\ref{fig:regime_4}a), but not necessarily (see figure~\ref{fig:regime_4}b).

\end{itemize}

% ------------------------------------------------------------- regime map -----

\subsection{Regime map and boundaries of the \textit{drops in jet} regime}

In view of using the \textit{drops in jet} regime for new applications such as the production of advanced fibers, it is required to describe the conditions of its occurrence. 
Since the full description of such collisions involves 13 independent parameters, we aim to focus on the most critical contributions and propose an analysis based on our experimental observations, studies of immiscible drop collisions, and knowledge about jet stability. Thus, two different fragmentation mechanisms of the continuous jet can be identified.

\begin{figure}[h]
\centering
\input{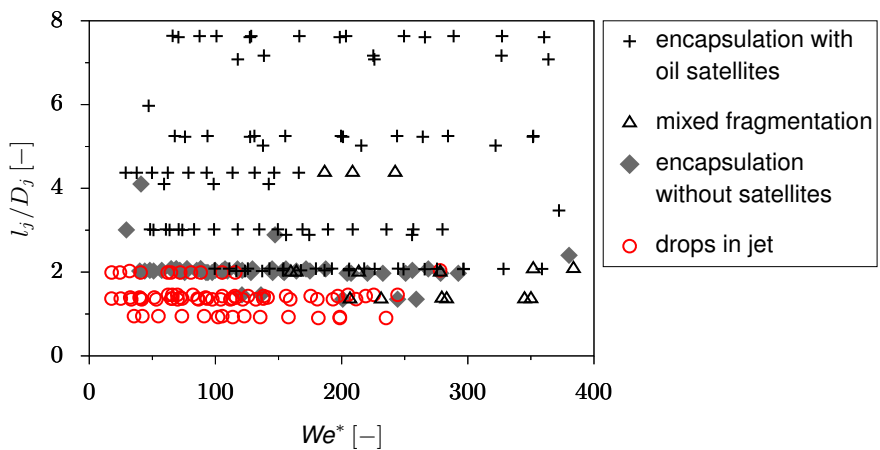}

\caption{Two-dimensional regime map. Using $l_j / D_j$ and $\Web^*$ allows to describe the boundaries of the \textit{drops in jet} regime. }
    \label{fig:regime_map}
\end{figure}

First, the jet fragmentation can be driven by inertia, as also observed for immiscible drop collisions. More precisely, this may occur if the kinetic energy of the impacting drops is too high by comparison to surface energy and viscous losses opposing jet distortion. In this case, it is expected that the drops simply cross over the jet - similarly to the crossing separation observed for immiscible drop collisions \cite{ref:Planchette-Lorenceau-Brenn_2012} - and fragment it. Based on this picture, and neglecting for simplicity the viscous losses, a first dimensionless parameter can be proposed that should help distinguishing the \textit{drops in jet} regime from the others. The impacting drop kinetic energy can be estimated by $\rho_d {D_d}^3 U^2$, while the surface energy opposing the crossing of the jet scales as $\sigma_j D_j l_j$. The ratio of these two quantities provides a modified Weber number defined by $\Web^*=\rho_d ({D_d}^3/D_j l_j) U^2/\sigma_j $. Using this parameter as the abscissa of the two-dimensional regime map of figure \ref{fig:regime_map}, we observe a good representation of the experimental results. The threshold value of $\Web^*$ is found to be approximately $240$. Above this value, no \textit{drops in jet} can be observed any more. Instead, mixed fragmentation becomes dominating in agreement with an inertial regime.

A second mechanism of fragmentation can be observed which corresponds to a Plateau-Rayleigh like instability where the continuous jet is destabilized by the periodic disturbances caused by the impacting drops. For jets of one liquid only, it is well known that the disturbances are unstable as soon as their wavelength $\lambda$ reaches the critical value of $\pi D_j$ \cite{ref:Rayleigh}. The Rayleigh criterion is often formulated as $\lambda/D_j \geq \pi$. By extension, it appears relevant for the studied collisions to consider the ratio between the drop spacing after contact with the jet and the jet diameter.  By definition, the  drop spacing after contact with the jet corresponds to $l_j$, the spacial period of the jet. It yields $l_j / D_j \geq \pi$. 

\begin{figure}[h]
\centering
  \parbox[c]{0.03\textwidth}{(a)}
  \hspace*{0mm}
  \parbox[c]{0.75\textwidth}{\includegraphics[angle=90,width=0.82\textwidth]{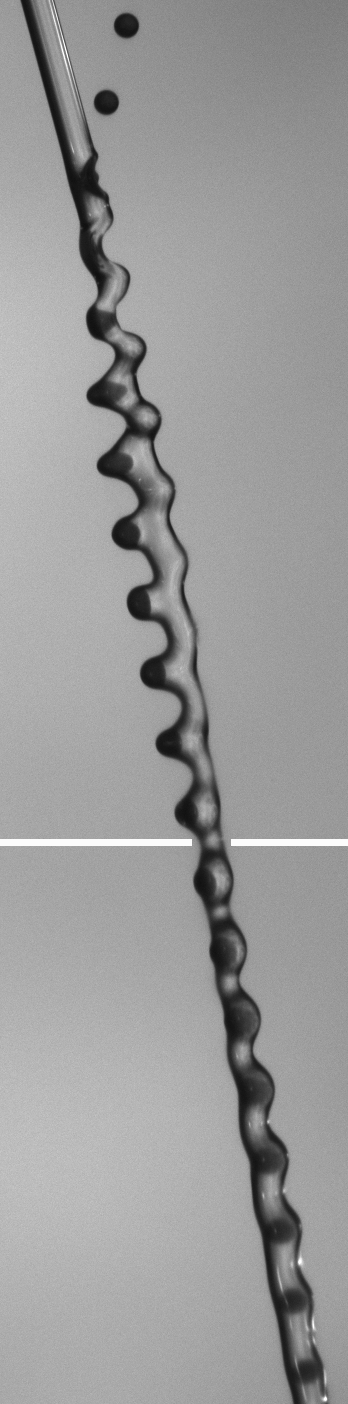}}\\
   \vspace{2mm}
  \parbox[c]{0.03\textwidth}{(b)}
  \hspace*{0mm}
  \parbox[c]{0.75\textwidth}{\includegraphics[angle=90,width=0.82\textwidth]{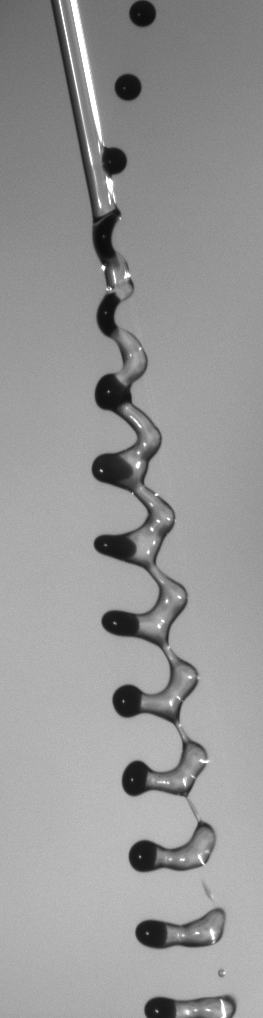}}\\
   \vspace{2mm}
	\parbox[c]{0.03\textwidth}{(c)}
	 \hspace*{0mm}
	\parbox[c]{0.75\textwidth}{\includegraphics[angle=90,width=0.82\textwidth]{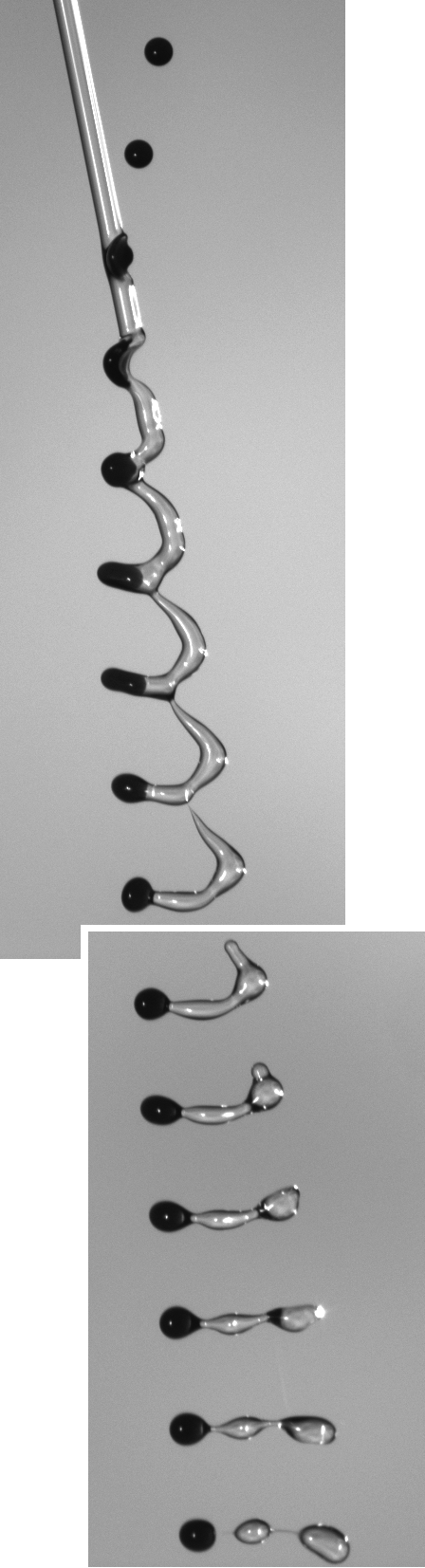}}\\
\caption{
(a)~$D_{d}=\SI{216}{\micro\meter}$, $D_{j}=\SI{300}{\micro\meter}$,
    $u_{d}=\SI{5.55}{\meter\per\second}$, $u_{j}=\SI{4.92}{\meter\per\second}$, $U=\SI{2.73}{\meter\per\second}$ yielding $l_j/D_j=1.99$ and $\Web^*=24$;
(b)~$D_{d}=\SI{202}{\micro\meter}$, $D_{j}=\SI{200}{\micro\meter}$,
    $u_{d}=\SI{6.73}{\meter\per\second}$, $u_{j}=\SI{4.90}{\meter\per\second}$, $U=\SI{3.15}{\meter\per\second}$ yielding $l_j/D_j=2.80$ and $\Web^*=34$ and (c)~$D_{d}=\SI{225}{\micro\meter}$, $D_{j}=\SI{200}{\micro\meter}$,
    $u_{d}=\SI{6.67}{\meter\per\second}$, $u_{j}=\SI{7.21}{\meter\per\second}$, $U=\SI{2.77}{\meter\per\second}$ yielding $l_j/D_j=4.04$ and $\Web^*=29$.}
    \label{fig:rayleigh_breakup}
\end{figure}

The use of $l_j / D_j $ as the ordinate variable of the two-dimensional regime map of figure \ref{fig:regime_map} confirms that $l_j / D_j $ is a critical parameter for the occurrence of \textit{drops in jet}. Yet, the threshold value is found to be approximately 2, \SI{50}{\percent} below the expected value of $\pi$, and seems to slightly decrease with the modified Weber number. This discrepancy can be explained by observing the typical pictures of figure \ref{fig:rayleigh_breakup}. 
The pictures of figure~\ref{fig:rayleigh_breakup} correspond to low $\Web^{*}$ and have been chosen to illustrate collisions where $l_{j}/D_{j}<2$; $2<l_{j}/D_{j}<\pi$ and $l_{j}/D_{j}>\pi$.
While theoretically both collisions with $l_{j}/D_{j}<\pi$ should lead to \textit{drops in jet}, only the one with $l_{j}/D_{j}<2$ does. We attribute this shift to the distortion of the jet subjected to the drop impacts. More precisely, the jet portion located between two impacting drops tends to be elongated by the relative movements of the drops and the jet whose relative momentum, normal to the final \textit{drops in jet} trajectory, is not yet fully dissipated. From $\Web^{*}$ definition, we expect that the greater $\Web^{*}$ is, the longer and thinner this cylindrical portion becomes. Thus, the underestimation of its aspect ratio when using $l_j / D_j $, e.g. the undistorted dimensions, increases with increasing $\Web^{*}$.

% ------------------------------------------------------------------------------
% ------------------------------------------------------------ conclusions -----

\section{Conclusions}

By colliding a drop stream with a continuous immiscible liquid jet we generate different kinds of liquid structures. These structures have been classified in 4 distinct regimes: \textit{drops in jet} for which both initial structures combine without any fragmentation; two regimes of encapsulation, both leading to the coating of the initial drops by a liquid shell originating from the jet.
The regimes can be distinguished by the formation (or not) of satellites and referred to as \textit{encapsulation with and without oil satellites}.
Finally, \textit{mixed fragmentation} occurs, where both the drops and the jet fragment into several droplets of various compositions.
To our knowledge, it is the first time that the formation of the very regular \textit{drops in jet} structure is reported. Two critical dimensionless parameters, corresponding to two distinct fragmentation mechanisms, have been proposed to describe the transitions between this regime and others. On the one hand, we identified a Plateau-Rayleigh like instability that results in the disappearance of the \textit{drops in jet} structure if $l_j / D_j $ reaches a critical value of approximately $2$. On the other hand, the fragmentation of the jet may also originate from an excess of kinetic energy. In this case, capillary forces are insufficient to keep the drops in the jet, and the drops fragment it by crossing over it. We derived a modified Weber number $\Web^*$  and verified that above a threshold value of approximately $240$, no \textit{drops in jet} structure can be obtained.

% ------------------------------------------------------------------------------
% ----------------------------------------------------------- nomenclature -----

\section {Nomenclature}
\begin{tabbing}
\hspace*{1.18 cm}  \= \hspace*{4 cm}  \kill
$D$           \> diameter \myunitbr{}{\si{\meter}} \\
$l$           \> spatial period length \myunitbr{}{\si{\meter}} \\
$f$           \> frequency of drop formation \myunitbr{}{\si{\hertz}} \\
$p$           \> impact parameter \myunitbr{}{\si{\meter}} \\
$\vec{u}, u$  \> velocity \myunitbr{}{\si{\meter\per\second}} \\
$\vec{U}, U$  \> relative velocity \myunitbr{}{\si{\meter\per\second}} \\
$\Web^*$        \> modified Weber number \\
$\vec{x}, \vec{y}, \vec{z}$   \> frame of reference \myunitbr{}{\si{\meter}} \\

$\alpha$  \> angle between jet and drop velocity vector \myunitbr{}{\si{\radian}} \\
$\beta$   \> angle between relative velocity vector and velocity vector \myunitbr{}{\si{\radian}} \\
$\lambda$ \> wavelength \myunitbr{}{\si{\meter}} \\
$\mu$     \> dynamic viscosity \myunitbr{}{\si{\pascal\second}} \\
$\rho$    \> density \myunitbr{}{\si{\kilogram\per\cubic\metre}} \\
$\sigma$  \> surface or interfacial tension \myunitbr{}{\si{\newton\per\metre}} \\
\end{tabbing}

\section {Subscripts}
\begin{tabbing}
\hspace*{1.18 cm}  \= \hspace*{4 cm}  \kill
$d$ \> drop \\
$j$ \> jet \\
\end{tabbing}

% ------------------------------------------------------------------------------
% ------------------------------------------------------------- references -----

\end{document}